\documentclass[sn-mathphys-num]{sn-jnl}

\usepackage{graphicx}
\usepackage{multirow}
\usepackage{amsmath,amssymb,amsfonts}
\usepackage{amsthm}
\usepackage{mathrsfs}
\usepackage[title]{appendix}
\usepackage{xcolor}
\usepackage{textcomp}
\usepackage{manyfoot}
\usepackage{booktabs}
\usepackage{algorithm}
\usepackage{upgreek}
\usepackage{algorithmicx}
\usepackage{algpseudocode}
\usepackage{listings}
\usepackage[normalem]{ulem}
\usepackage{braket}

\begin{document}

\title[Article Title]{Direct experimental access to the bulk band inversion in a topological metamaterial}



\author*[1,2]{\fnm{Simon} \sur{Widmann}}\email{simon.widmann@uni-wuerzburg.de}
\author[1,2]{\fnm{Johannes} \sur{Düreth}}
\author[1,2]{\fnm{Siddhartha} \sur{Dam}}
\author[1,2]{\fnm{Christian G.} \sur{Mayer}}
\author[1,2]{\fnm{David} \sur{Laibacher}}
\author[1,2]{\fnm{Monika} \sur{Emmerling}}
\author[1,2]{\fnm{Martin} \sur{Kamp}}

\author[2,3]{\fnm{Friedrich} \sur{Reinert}}
\author[2,3]{\fnm{Maximilian} \sur{Ünzelmann}}
\author[1,2]{\fnm{Simon} \sur{Betzold}}
\author[1,2]{\fnm{Sven} \sur{Höfling}}
\author*[1,2]{\fnm{Sebastian} \sur{Klembt}}\email{sebastian.klembt@uni-wuerzburg.de}

\affil[1]{\orgdiv{Technische Physik}, \orgname{Universität Würzburg}, \orgaddress{\street{Am Hubland}, \city{Würzburg}, \postcode{97074}, \country{Germany}}}
\affil[2]{\orgdiv{Würzburg-Dresden Cluster of Excellence ct.qmat}, \orgname{Universität Würzburg}, \orgaddress{\street{Am Hubland}, \city{Würzburg}, \postcode{97074}, \country{Germany}}}
\affil[3]{\orgdiv{Lehrstuhl für Experimentelle Physik VII}, \orgname{Universität Würzburg}, \orgaddress{\street{Am Hubland}, \city{Würzburg}, \postcode{97074}, \country{Germany}}}

\abstract{
Topological phases in exciton-polaritons and other metamaterial platforms have attracted significant attention due to their flexibility as Hamiltonian simulators. 
In previous works, signatures of topology have mainly been investigated from the perspective of edge states -- strongly localised modes with exponentially decaying intensity into the bulk. 
While these edge states have become the hallmark of topological systems as they can facilitate non-reciprocal transport in potential applications, the topology is fundamentally encoded in the bulk band structure.
In particular, the momentum-dependence of the eigenstates, i.e., the wave functions, determines the topology, usually reflected in a bulk band inversion.
We present a band inversion in the paradigmatic Su-Schrieffer-Heeger (SSH) model, characterised by a reversal of the sublattice symmetry, quantified by the expectation value $\langle \sigma_\mathrm{x} \rangle$, when going from the centre of the Brillouin zone to the zone boundary.
Here, we show direct experimental access to this bulk band inversion in SSH exciton-polariton chains, using two-dimensional momentum-space ($k$-space) mapping -- without the need for real-space imaging.
This technique enables the direct observation of the momentum-dependent inversion of the sublattice symmetry in the bulk bands, providing a unique perspective on topological phases beyond conventional edge state measurements. 
Our approach establishes effective momentum-resolved sublattice phase measurements as a powerful tool for accessing the wave function and bulk topology in photonic and polaritonic systems and beyond.
}
\keywords{Topological photonics, Su-Schrieffer-Heeger (SSH), Topological band inversion, Bulk boundary correspondence, Exciton-Polariton, Hamiltonian Simulators}

\maketitle


Topological insulators are materials that behave as electrical insulators in their bulk but support conducting states at their surfaces or edges~\cite{Bernevig2006,Konig2007}.
These boundary states arise from the material’s nontrivial topological order and are protected by symmetries, making them robust against disorder and imperfections.
The discovery of topological insulators has opened new directions in condensed matter physics, linking quantum mechanics and materials science~\cite{Hasan2010, Shankar22}.
Their unique properties hold promise for applications in spintronics~\cite{Pesin2012} and quantum computing~\cite{Hasan2010, Breunig2022}.
More recently, the field of topological photonics has emerged~\cite{Rechtsman2013, Hafezi2013}, where analogous principles enable robust light transport in engineered photonic structures.

Experimentally detecting the topological interface modes is a key technique to confirm the presence of topological phases, as their existence directly reflects the bulk-boundary correspondence principle: The nontrivial topology of the bulk guarantees robust, gapless modes at the interfaces where the topological invariant changes.
In electronic systems, these modes can be experimentally investigated using real-space techniques such as scanning probe measurements~\cite{Setescak25, Erhardt25, Reis17, Hsieh2009_Science} or momentum-space techniques like angle-resolved photoemission spectroscopy~\cite{Hsieh2009_Nature}.
In photonic metamaterials, topological interface states are generally accessible due to the larger lattice constants using real-space spectroscopic methods~\cite{Slobozhanyuk19, Yan21, Rechtsman2013, Hafezi2013, Bandres18}, but also using momentum-space spectroscopy~\cite{Klembt2018, Li2023, Pickup2020, St-Jean2021, Su2025, Guillot2025}.
While a nontrivial bulk topology is often inferred from the observation of exponentially localised boundary modes, these modes are merely a result of a topological bulk band inversion.

Lately, there is a growing interest in developing innovative methods to directly access bulk topological properties~\cite{Che2020,Su2025,St-Jean2021,Guillot2025}.
These approaches comprise imaging bulk modes in real and momentum-space including phase information, measuring the response of the system to perturbations, or analysing the dynamics of wavepacket propagation to reveal features like Berry curvature, Chern numbers, or Zak phases~\cite{Li2023, St-Jean2021, Guillot2025, Lackner25}.
The goal is to establish a clear connection between the observable properties and the underlying topological order of the system.
This not only enables the identification and characterisation of topological phases, but also guides the design of robust devices that exploit topologically protected interface states.

In this work, we demonstrate direct experimental access to the bulk topology in Su-Schrieffer-Heeger (SSH) exciton-polariton chains solely requiring momentum-space spectroscopy~\cite{Su1980}.
By displacing the sublattices perpendicular to the chain, we control the effective sublattice phase shifts, enabling direct observation of the momentum-dependent inversion of sublattice symmetry in the bulk bands.
This approach allows probing topological phases beyond conventional edge-state measurements, establishing sublattice-phase-resolved $k$-space spectroscopy as a versatile tool for accessing bulk wave functions and topology.

We utilise a III-V semiconductor platform for the implementation of the SSH chains.
Our system hosts exciton-polaritons, hybrid light-matter quasiparticles formed by the strong coupling of photons and excitons in semiconductor quantum well microcavities~\cite{Weisbuch1992, Deng10}.
Embedding the quantum wells in a high-quality Fabry-Pérot microcavity enables this strong coupling, producing new eigenmodes known as the upper and lower polariton branches.
Exciton-polaritons provide an ideal platform for topological photonics and Hamiltonian emulation due to their hybrid light-matter nature~\cite{Amo2016,Schneider2016, Klembt2018, Lu2014, Ozawa2019}.
Their band structures and wave functions can be directly probed via real- and momentum-space spectroscopy, while flexible lattice engineering allows precise implementation of complex Hamiltonians with nontrivial topology.
Additionally, their driven-dissipative dynamics make them suitable for exploring non-Hermitian and out-of-equilibrium topological phenomena, offering a versatile platform for studying both fundamental physics and potential photonic devices.
Thus, polariton lattices provide a versatile platform for exploring topological photonics, enabling the realisation of models ranging from SSH chains~\cite{St-Jean2017, Harder2021, Pickup20, Dusel2021, Gagel2024, Lackner25, Georgakilas2025}, to Chern insulators~\cite{Klembt2018, Widmann2025}, 2D SSH lattices hosting higher-order topological states~\cite{Wu2023} and others~\cite{Liu2020,Li2021}.

\section*{SSH bulk band inversion}
\begin{figure}[h!]
\centering
\includegraphics[width=0.9\textwidth]{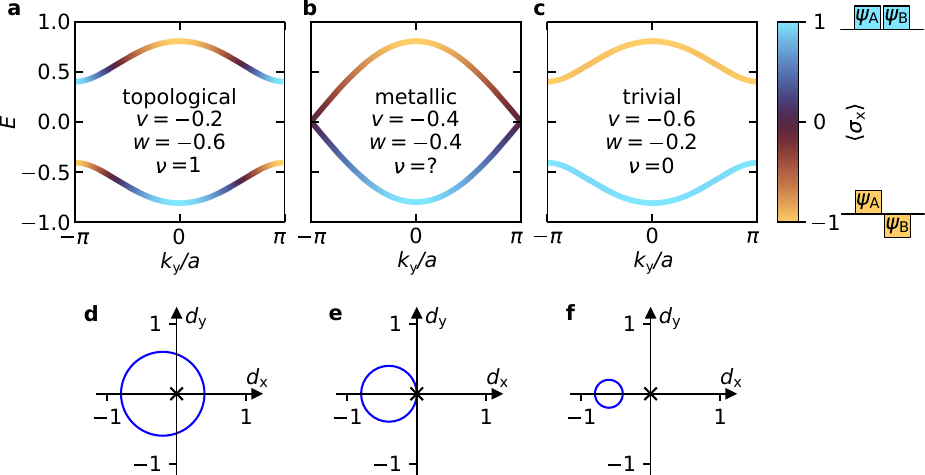}
\caption{
\textbf{Band inversion and parametrisation of the Hamiltonian. a-c}, Band structures of the one-dimensional Su-Schrieffer-Heeger (SSH) Hamiltonian for different intra and inter-cell hoppings $v$ and $w$. The different chains are in the topological (\textbf{a}), metallic (\textbf{b}) and trivial insulator phase (\textbf{c}). The bands are overlaid with the sublattice symmetry $\langle \sigma_\mathrm{x} \rangle$ that distinctly inverts at the edge of the Brillouin zone in the topological case (\textbf{a}). \textbf{d-f}, Parametrisation of the Hamiltonian in the Pauli matrix basis $\hat{H}(k_\mathrm{y})=d_\mathrm{x}(k_\mathrm{y})\sigma_\mathrm{x}+d_\mathrm{y}(k_\mathrm{y})\sigma_\mathrm{y}$ when passing through the Brillouin zone. The winding number $\nu$ of this circle is the topological invariant of the SSH model~\cite{Asboth2016}.
}
\label{fig:theory}
\end{figure}

The Bloch Hamiltonian of the SSH model with intra-cell hopping $v$ and inter-cell hopping $w$ is given by
\begin{align}
    \hat{H}(k_\mathrm{y}) = \begin{pmatrix}
0 & v + w e^{-ik_\mathrm{y} a} \\
v + w e^{ik_\mathrm{y} a} & 0
\end{pmatrix} \;\;.
\label{eq:hamiltonian}
\end{align}
where $a$ is the lattice constant (unit cell length) and $k_\mathrm{y}$ is the wave vector~\cite{Asboth2016}.
For consistency with the experimental part, $k_\mathrm{y}$ is used as the wave vector parallel to the chain direction.
Due to $\hat{H}(k_\mathrm{y})^2 \vec{\psi}\pm =E(k_\mathrm{y})^2 \hat{\mathbb{I}}_{2\times2}\vec{\psi}\pm$, the lower and upper band of $\hat{H}(k_\mathrm{y})$ are given by $E_\pm(k_\mathrm{y})=\pm\sqrt{v^2 + w^2 + 2 v w \cos(k_\mathrm{y}a)}$ with the corresponding eigenstates $\vec{\psi}\pm=(\psi_\mathrm{A\pm}, \psi_\mathrm{B\pm})^T$.
The band structures are displayed in Fig.\,\ref{fig:theory}a-c, where the intra and inter-cell hoppings are varied with $|v|<|w|$ in a, $v=w$ in b and $|v|>|w|$ in \textbf{c}.

Rewriting the Bloch Hamiltonian in a Pauli matrix basis $H(k_\mathrm{y})=d_\mathrm{x}(k_\mathrm{y}) \sigma_\mathrm{x} +d_\mathrm{y}(k_\mathrm{y}) \sigma_\mathrm{y} +d_\mathrm{z}(k_\mathrm{y}) \sigma_\mathrm{z} + d_\mathrm{0}(k_\mathrm{y}) \hat{\mathbb{I}}_{2\times2}$ yields $d_\mathrm{x}(k_\mathrm{y})=v + w \cos(k_\mathrm{y}a)$, $d_\mathrm{y}(k_\mathrm{y})=w \sin(k_\mathrm{y}a)$ and $d_\mathrm{0}(k_\mathrm{y}) = d_\mathrm{z}(k_\mathrm{y}) = 0$.
As displayed in Fig.\,\ref{fig:theory}d-f, the only non-zero contributions $d_\mathrm{x}(k_\mathrm{y})$ and $d_\mathrm{y}(k_\mathrm{y})$ parametrise a circle with radius $|w|$ and an offset of $v$ on the $d_\mathrm{x}$-axis.
\\
Every topological insulator is classified by a topological invariant, an integer number that does not change under adiabatic transformations which do not close and reopen the band gap and preserve the underlying symmetry~\cite{Hasan2010}.
In the case of the SSH model, this invariant is given by the winding number $\nu$ of the closed curve parametrised in the $d_\mathrm{x}$--$d_\mathrm{y}$--plane for parameters $k_\mathrm{y}\in[-\frac{\pi}{a},\frac{\pi}{a}]$.
Intuitively, the winding number counts how many times this circle wraps around the origin of the $d_\mathrm{x}$--$d_\mathrm{y}$--coordinate system.
In the case of $\nu=1$ (topologically non-trivial case), the origin is included, while $\nu=0$ (topologically trivial case) means that the circle does not include the origin~\cite{Asboth2016}.
The parametrised circles in Fig.\,\ref{fig:theory}d-f for the different combinations of hopping above behave accordingly.
Evidently, the SSH model is in the topologically non-trivial phase, when $|v|<|w|$ (Fig.\,\ref{fig:theory}a and d), while $|v|>|w|$ (Fig.\,\ref{fig:theory}c and f) yields a trivial insulator.
In the metallic phase with $v=w$ (Fig.\,\ref{fig:theory}b and e), the winding number is not defined and the band gap closes at the edges of the Brillouin zone $k_\mathrm{y}=\pm \frac{\pi}{a}$.
The topologically non-trivial phase gives rise to in-gap edge states in a finite chain in real-space. 

While these edge states are abundantly reported, they are rather the response to the bulk band topology, reflected by the band inversion.
We now show that the occurrence of these topological in-gap edge states is a direct response to a bulk band inversion at the edge of the Brillouin zone.
In the case of the SSH model, this band inversion occurs in the expectation value of the first Pauli matrix, i.e. 
\begin{align}
 \langle \sigma_\mathrm{x} \rangle &=
 \begin{pmatrix}
\psi_\mathrm{A}^* & \psi_\mathrm{B}^* 
\end{pmatrix}
\begin{pmatrix}
0 & 1 \\
1 & 0
\end{pmatrix}
\begin{pmatrix}
\psi_\mathrm{A} \\
\psi_\mathrm{B} 
\end{pmatrix}
=
\begin{pmatrix}
\psi_\mathrm{A}^* & \psi_\mathrm{B}^* 
\end{pmatrix}
\begin{pmatrix}
\psi_\mathrm{B} \\
\psi_\mathrm{A} 
\end{pmatrix}\\
&=\psi_\mathrm{A}^*\psi_\mathrm{B}+\psi_\mathrm{B}^* \psi_\mathrm{A}\;.
\label{eq:sigmaXBandInv}
\end{align}
Intuitively, it probes the phase between the sublattice entries for a given eigenvector, namely $\psi_\mathrm{A}$ and $\psi_\mathrm{B}$.
Here, $\langle \sigma_\mathrm{x} \rangle=+1$ corresponds to zero phase shift $\Delta \theta = 0$ between the sublattices, while $\langle \sigma_\mathrm{x} \rangle=-1$ equates to a phase shift of $\Delta \theta = \pi$, as depicted in the colour bar in Fig.\,\ref{fig:theory}c.
\\
This sublattice symmetry is overlaid on the bands in Fig.\,\ref{fig:theory}a-c for the topological, metallic and trivial case.
While each band has constant sublattice symmetry in the trivial case, the sublattice symmetry inverts at the edge of the Brillouin zone in the non-trivial case -- i.e. a band inversion occurs as a direct signature of the bulk topology -- without the presence of topological edge states.

\section*{Real-space access: topological edge states}
\begin{figure}[h!]
\centering
\includegraphics[width=0.8\textwidth]{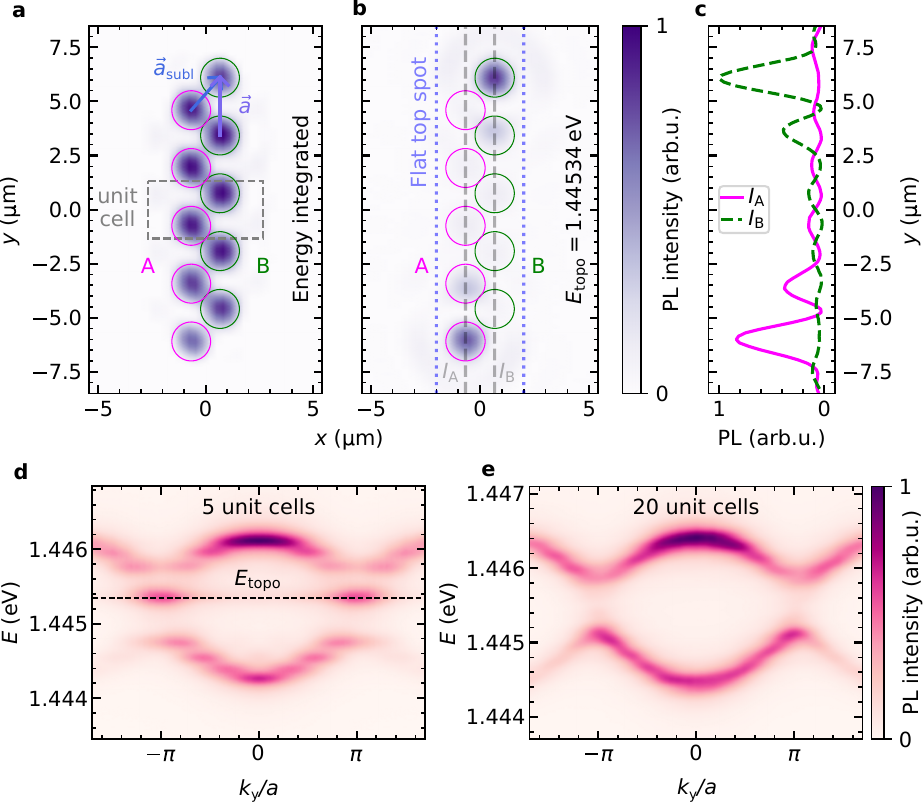}
\caption{   
\textbf{Real and reciprocal space signatures of topology. a}, Real-space distribution of a topologically non--trivial polariton SSH chain, energy integrated over the full $s$--band. The chain is oriented in $y$--direction, while the sublattices $A$ and $B$ are displaced in $x$--direction. \textbf{b}, Constant energy cut through the topological edge states of the chain. \textbf{c}, Cuts along the dashed grey lines through the two sublattices indicated in \textbf{b}. The intensity in the lattice sites decays when moving away from the respective edge. \textbf{d}, Band structure of the chain shown in \textbf{a} with visible discretisation. The edge state is visible in the band gap at $E_\mathrm{topo}$. \textbf{e}, Band structure of a chain consisting of 20 unit cells.
}
\label{fig:classic_polariton}
\end{figure}

In order to experimentally investigate the bulk topology of the SSH model, we implement exciton-polariton chains with different hopping configurations using the etch-and-overgrowth molecular beam epitaxy technique~\cite{Kaitouni06}.
Using this technique, we locally elongate the cavity, tuning the photonic potential landscape.
More details on the growth and patterning process are given in the Methods.
To study the band structures using Fourier space spectroscopy while adding an extra effective phase shift to the light that originates from the two sublattices, we realise exciton-polariton chains where the sublattices are shifted in real-space perpendicular to the chain direction~\cite{Harder2021}.
More specifically, we start with a zig-zag chain with circular mesas of diameter $d=1.8\,\mathrm{\upmu m}$ and a lattice constant of $a=2\cdot1.05\cdot d=3.78\,\mathrm{\upmu m}$ in $y$-direction, where sublattice $A$ is shifted by $a/2$ in negative $x$-direction relative to sublattice $B$.
At the negative $y$-end, the chain starts with sublattice $A$.
A sketch of the layout with the two sublattices is shown in Fig.\,\ref{fig:classic_polariton}a.
In order to implement the staggered hopping $v\neq w$, the two sublattices are additionally shifted slightly in opposing directions in $y$.
The chain in Fig.\,\ref{fig:classic_polariton}a realises a lower intra-cell hopping $v$ compared to the inter-cell hopping $w$ (topological configuration) by displacing $A$ ($B$) in negative (positive) $y$-direction by $\delta y=84\,\mathrm{nm}$.
The resulting vector that connects the sublattices is thus given by
$\vec{a}_\mathrm{subl}=\left(a/2, a/2-2 \delta y \right)$.
\\
All polariton structures studied in this work are excited off-resonantly in the first high-energy Bragg minimum of the microcavity.
The continuous wave laser spot is shaped into a rectangular flattop with a width of approximately $\Delta x \approx 4\,\mathrm{\upmu m}$ (vertical dashed lines in Fig.\,\ref{fig:classic_polariton}b) and a length $\Delta y \approx 60\,\mathrm{\upmu m}$ using a spatial light modulator.
This way, the contribution of individual eigenstates to the far-field does not depend on their localisation in real-space.
This is underlined by the spatially homogeneous photoluminescence shown in Fig.\,\ref{fig:classic_polariton}a, which homogeneously originates from all lattice sites.
The colour map plot in Fig.\,\ref{fig:classic_polariton}a shows the $s$-band energy integrated photoluminescence of the topological chain with a length of five unit cells.
This measurement is taken by scanning a real-space image of the sample photoluminescence over the spectrometer entrance slit, taking $E$ over $y$ spectra for a range of $x$-positions.
\\
While Fig.\,\ref{fig:classic_polariton}a is energy integrated, a constant energy cut at $E_\mathrm{topo}=1.44534\,\mathrm{eV}$ reveals the topological edge states in Fig.\,\ref{fig:classic_polariton}b, which are located at the edges of the respective sublattices as shown by the sublattice centre cuts $I_\mathrm{A}$ and $I_\mathrm{B}$ in Fig.\,\ref{fig:classic_polariton}c.
The edge states clearly appear as in-gap states in the band structure of the chain as depicted in Fig.\,\ref{fig:classic_polariton}d.
This band structure exhibits visible discretisation due to the finite chain length of 5 unit cells.
While the band structure of the chain is only dispersive along $k_\mathrm{y}$, the full two-dimensional far-field of the system is measured in order to reconstruct the band inversion as explained below.
The band structures shown in Fig.\,\ref{fig:classic_polariton}d and e are integrated along $k_\mathrm{x}$, in order to make the upper band visible at $k_\mathrm{y}=0$, which is otherwise absent due to destructive interference (structure factor).
While discretisation is present in a 5 unit cell chain, the bands appear homogeneous when going to a chain that comprises 20 unit cells as shown in Fig.\,\ref{fig:classic_polariton}e.
While all other experimental parameters of this chain are identical to the short chain presented in Fig.\,\ref{fig:classic_polariton}d, the eigenstates of the longer chain are slightly shifted in energy because of the different position of the chains on the sample, which has an inherent detuning gradient due to the monolithic growth process. 
The relative intensity of the edge states in the band structure of the longer chain decreases, as the number of bulk modes grows linearly with the number of unit cells.
However, only two edge states are present, regardless of chain length.

\section*{Reciprocal space access: topological band inversion}
In order to reconstruct the sublattice symmetry $\langle \sigma_\mathrm{x} \rangle \propto I_0 - I_\pi$ (see Methods for the mathematical derivation), the far-field has to be measured for the effective sublattice phase shifts $\Delta \theta = 0$ and $\Delta \theta = \pi$.
These can be accessed experimentally by probing the far-field along a line that is perpendicular to the vector $\vec{a}_\mathrm{subl}=\left(a/2, a/2-2 \delta y \right)$ that connects the sublattices.
While the zero effective sublattice phase shift line ($\Delta \theta = 0$) includes the origin $\vec{k}=\vec{0}$, the $\pi$ phase shift line is also perpendicular to $\vec{a}_\mathrm{subl}$, but offset parallel to $\vec{a}_\mathrm{subl}$ by
\begin{align*}
\Delta \vec{k}_\pi = \frac{\pi}{|\vec{a}_\mathrm{subl}|}\cdot \frac{\vec{a}_\mathrm{subl}}{|\vec{a}_\mathrm{subl}|}
\end{align*}
as shown in the Methods.
Although the chain is not periodic in this direction, in the realm of solid-state physics, $\Delta \vec{k}_\pi$ would equate to the edge of the first Brillouin (i.e. $\pi$ phase shift) in the sublattice direction.
While the presented physics is already observable in the two far-field cuts explained above, we utilise the full $k$-space accessible with the $0.65$ numerical aperture ($\mathrm{NA}$) objective and overlay all equivalent cuts that correspond to $0$ and $\pi$ effective sublattice phase shifts as explained in the Methods.
The individual cuts through the raw data of the topological chain are shown in the Methods in Fig.\,\ref{fig:methods:indivCuts}.
While the use of a large NA objective is necessary in order to access higher $|\vec{k}|$ effective sublattice phase shift cuts, this automatically corresponds to a relatively high magnification ($50\times$), limiting the area in real-space that can be probed experimentally.
The need for spatially constant excitation limits the maximum accessible chain size to roughly 20 unit cells.
If parts of the chain were not excited, the excitation of different eigenmodes would vary depending on their spatial localisation, influencing the band structure intensity and thus our evaluation that relies on relative band intensity.

\begin{figure}[h!]
\centering
\includegraphics[width=1\textwidth]{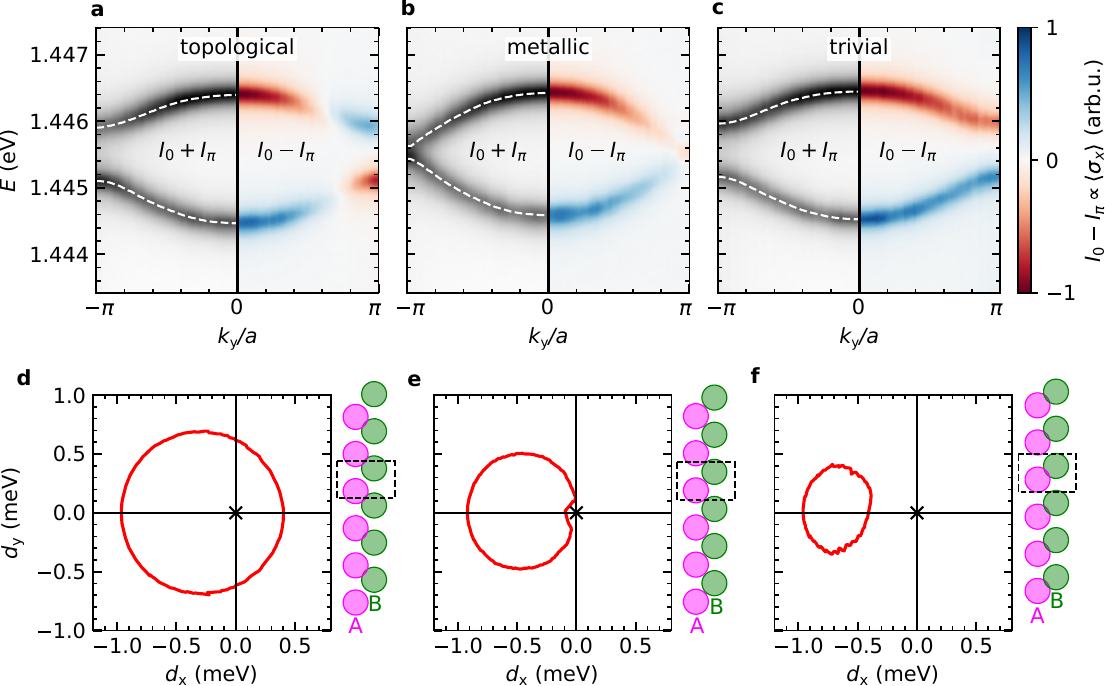}
\caption{   
\textbf{Experimental access to the bulk band inversion. a-c}, Band structures and difference spectra for $\Delta \theta = 0$ and $\Delta \theta = \pi$ of chains comprising 20 unit cells. The chains are in a topological (a), metallic (b) and trivial configuration (c). The experimental band inversion can be seen for the topological chain. \textbf{d-f}, Reconstructed Hamiltonian in a Pauli matrix basis from the sublattice phase fitting procedure. The closed curves define the winding number. For $|v|<|w|$, the curve contains the origin of the $d_\mathrm{x}$--$d_\mathrm{y}$ coordinate system, hinting to nontrivial topology. For each of the chain configurations, the real-space location of the mesas is sketched. The sublattice relative $y$-shift of the gapped chains is exaggerated for illustrative purposes.
}
\label{fig:bandInversion}
\end{figure}

Far-field band structures are generally subject to sublattice interference when a Bloch state comprises more than one orbital, due to the structure factor.
Our method allows for determining the actual structure-factor-corrected far-field by overlaying the two individual $0$ and $\pi$ effective sublattice phase shift far-field spectra $I_0+I_\pi$, as displayed on the left sides of Fig.\,\ref{fig:bandInversion}a-c for negative $k_\mathrm{y}$, resulting in nearly constant band intensity.
The band structures for the topological (Fig.\,\ref{fig:bandInversion}a) and trivial insulator chain (Fig.\,\ref{fig:bandInversion}c) appear identical besides a slight energy difference due to the detuning gradient of the sample as explained above.
The chains presented have a sublattice displacement of $\delta y={84\,\mathrm{nm}, 0\,\mathrm{nm}, -84\,\mathrm{nm}}$ from left to right, where $|\delta y|=84\,\mathrm{nm}$ is on the order of $4\%$ of the spacing between lattice sites.
Notably, the $\delta y=84\,\mathrm{nm}$ and $\delta y=-84\,\mathrm{nm}$ chains are identical up to their edge termination.
\\
Remarkably, in contrast to the constant intensity band structures, the sublattice symmetry of the bulk bands shows a band inversion at the edge of the Brillouin zone in the difference spectrum $I_0 - I_\pi\propto\langle \sigma_\mathrm{x}\rangle$ of the topological SSH chain (Fig.\,\ref{fig:bandInversion}a), while revealing a constant distribution of the sublattice symmetry for the two bands in the trivial case.
Therein, the lower energy band has eigenstates with $0$ phase shift, while the upper band eigenstates have $\pi$ phase shift.
In the metallic case (Fig.\,\ref{fig:bandInversion}b), the sublattice symmetry vanishes at $k_\mathrm{y}/a=\pm\pi$.
The experimental observations align with the theoretical sublattice symmetry calculated for the tight binding Bloch Hamiltonian in Fig.\,\ref{fig:theory}a-c, demonstrating the band inversion in the sublattice symmetry in the topologically non-trivial case.
\\
\section*{Hamiltonian reconstruction}
While the band inversion already conveys the system topology, utilising the full effective sublattice phase (arbitrary $\Delta \theta$) dependent far-field spectra allows for a direct reconstruction of the systems effective $2\times2$ Hamiltonian.
In order to reconstruct the Hamiltonian with eigenenergies $E_-$ and $E_+$ in the framework of a tight binding model, it is necessary to determine the corresponding eigenstates $\ket{\psi_-}=(\psi_{\mathrm{A},-}, \psi_{\mathrm{B},-})^T$ and $\ket{\psi_+}=(\psi_{\mathrm{A},+}, \psi_{\mathrm{B},+})^T$.
The Hamiltonian is then given in the spectral form $H=E_-\ket{\psi_-}\bra{\psi_-}+E_+\ket{\psi_+}\bra{\psi_+}$.
We therefore assume an eigenstate that is localised with equal probability of $0.5$ on both sublattices: $0.5=|\psi_{\mathrm{A},-}|^2=|\psi_{\mathrm{A},+}|^2=|\psi_{\mathrm{B},-}|^2=|\psi_{\mathrm{B},+}|^2$, as implied by the symmetries of the SSH model as well as our polariton confinement potential.
This assumption is underlined by the fact that the intensity goes to zero in the phase fitting procedure, which is detailed in the Methods.
The absolute phase of the wave function is gauge dependent $\ket{\psi_\pm}\rightarrow e^{i\alpha}\ket{\psi_\pm}$, hence the first entries of the two eigenstates are chosen $\frac{1}{\sqrt{2}}=\psi_{\mathrm{A},-}=\psi_{\mathrm{A},+}$.
The only unknown variable is thus the phase between the two sublattices, namely $\phi$.
This phase has to be determined for a given chain for each of the two bands $E_-$ and $E_+$ at each $k_\mathrm{y}$, as detailed in the Methods.

Analogous to the treatment of the SSH tight binding matrix (equation\,\ref{eq:hamiltonian}), the Hamiltonian can be expressed in the Pauli matrix basis.
Naturally, $d_\mathrm{z}$ (staggered on-site energy term) is zero due to the $50\,\%$ sublattice localisation assumption.
The Hamiltonians in the $d_\mathrm{x}$--$d_\mathrm{y}$--basis of the three investigated experimental configurations are shown in Fig.\,\ref{fig:bandInversion}d-f.
The Hamiltonian of the topological chain (Fig.\,\ref{fig:bandInversion}d) includes the origin of the $d_\mathrm{x}$--$d_\mathrm{y}$--coordinate system, while the trivial one (Fig.\,\ref{fig:bandInversion}f) does not.
In the metallic case, the curve parametrised by the Hamiltonian cuts through the origin, which is not completely reproduced by the experiment.
This deviation is rooted in the first step of the wave function fit: As evidenced from Fig.\,\ref{fig:bandInversion} b, the dual Lorentzian fit used to determine the band energies at each $k_\mathrm{y}$ does not find a fully closed gap. 
While the fitted Pauli basis Hamiltonian deviates more from a circular shape in the trivial case, it quantitatively exhibits the expected behaviour, encircling the origin in the non-trivial case and avoiding it in the trivial case.

\section*{Discussion and outlook}
In this work, we have presented a method to experimentally probe the bulk topology, revealing a bulk band inversion in $\langle \sigma_\mathrm{x} \rangle$. 
We have utilised an SSH exciton-polariton chain where the two sublattices are displaced perpendicular to the chain.
This allows for a direct experimental access of the sublattice symmetry $\langle \sigma_\mathrm{x} \rangle$, solely from far-field measurements, complementing other probing techniques such as the study of edge states or the Zak phase.
We show that the sublattice symmetry $\langle \sigma_\mathrm{x} \rangle$ inverts at the edge of the Brillouin zone for the topological chain, while the symmetry is strictly divided into a bonding and antibonding band in the trivial case.
While edge states are most ubiquitously addressed in topological metamaterials, our findings demonstrate how the topology is fundamentally encoded in the bulk eigenvectors of the photonic crystal.
In the Pauli basis, the Hamiltonian's $\sigma_\mathrm{x}$ and $\sigma_\mathrm{y}$ component parametrises the winding number -- the topological invariant of the SSH model.
We have reconstructed the effective $2\times2$ Hamiltonian of the system, underpinning the topologically distinct configurations of the presented geometries.
\\
While band topology has been studied extensively in other metamaterials, we present a method that purely relies on far-field measurements, offering transferability to techniques such as angle-resolved photoemission spectroscopy. 
In fact, recent studies have demonstrated that dichroic angle-resolved photoemission spectroscopy can provide access to topological pseudospin bulk band inversions in Weyl semimetals~\cite{Unzelmann2021} and quantum spin Hall insulators~\cite{Erhardt2024}.
Although band inversions have not been readily studied experimentally thus far, they are central to the concept of band topology and generally accessible, especially in metamaterials where the wave functions can be probed.
Specifically with regard to exciton-polariton lattices \cite{Klembt2018} or topological lasers \cite{Dikopoltsev2021, Peng24}, the technique can be transferred to two-dimensional systems by moving away from strict $k$-space methods, where sublattice phase shifts can be introduced by the use of a spatial light modulator in a real intermediate image plane~\cite{Guillot2025, Guillot2025_2}.

\bibliography{sn-bibliography}

\section*{Acknowledgements}
We acknowledge financial support by the German Research Foundation (DFG) under Germany’s Excellence Strategy-EXC2147 “ct.qmat” (project id 390858490).

\section*{Author contributions}
S.W., J.D., S.D. and C.G.M. built the experimental setup, performed the experiments and evaluated the data. S.D. grew the sample by molecular beam epitaxy.
M.E., S.W., D.L. and M.K. realised the layout and processing of the sample.
S.W., M.Ü., S.B. and F.R. performed the theoretical analysis.
S.H. and S.K. provided the funding.
S.W. and S.K. wrote the original draft of the manuscript with input from all authors.

\section*{Competing interests}
The authors declare no competing interests.

\section*{Data availability}
All datasets generated and analysed during this study are available upon reasonable request from the corresponding authors.




\newpage
\section*{Methods}\label{methods}
\subsection*{Sample growth}
The sample was grown using etch-and-overgrowth molecular beam epitaxy (MBE) to induce controlled, local variations of the cavity thickness~\cite{El-Daif2006,Winkler2015}. 
The epitaxial stack begins with a bottom distributed Bragg reflector (DBR) of $44.5$ pairs of $\mathrm{AlAs}/\mathrm{GaAs}$ layers with $\lambda/4$ optical thickness. 
On top, a $\lambda$-thick $\mathrm{GaAs}$ cavity was grown that embeds three $\mathrm{In_{0.06}Ga_{0.94}As}$ quantum wells positioned at the antinodes of the standing optical field, forming the optically active region.

For the confinement potential patterning, the wafer was taken ex situ, spin-coated with positive PMMA resist, and the lattice layout was written by electron-beam lithography. 
After development, a $20\,\mathrm{nm}$ aluminium film was thermally evaporated to define the hard mask. 
Lift-off was performed in pyrrolidone, and the exposed GaAs/AlAs layers were wet-etched in an $\mathrm{H_2O:H_2O_2}(30\%):\mathrm{H_2SO_4}(96\%)=(800:4:1)$ solution, with the etch depth controlled by time. 
The aluminium mask was then removed in a $1\%\,\mathrm{NaOH}$ bath, followed by a surface cleaning in $96\%\,\mathrm{H_2SO_4}$ for two minutes. 
Subsequently, the sample was reintroduced into the MBE chamber for overgrowth and surface oxides were removed using activated hydrogen. 
The microcavity was completed by depositing a top DBR consisting of 40 pairs of $\mathrm{AlAs/Al_{0.2}Ga_{0.8}As}$ quarter-wave layers.
The sample has a Rabi splitting of $(4.16\pm0.18)\,\mathrm{meV}$ and an exciton energy of $(1.4547\pm0.0010)\,\mathrm{eV}$ (at zero detuning).

\subsection*{Spectroscopy}
A wavelength-tunable continuous-wave laser (M-Squared SolsTis) was used to excite the microcavity off-resonantly in reflection geometry.
The pump beam was shaped into a rectangular flat-top profile with a Holoeye GAEA-2 spatial light modulator.
Excitation and collection were performed through the same microscope objective.
An infinity-corrected $50\times$ objective with a numerical aperture of $\mathrm{NA}=0.65$ (Mitutoyo Plan Apo NIR) was used, which allows access to large emission angles and thus to the higher-angle effective sublattice-phase cuts described in Methods~\ref{sec:kspacePath}.

The sample was mounted under vacuum in a continuous-flow liquid-helium cryostat (Janis ST-500) and temperature-stabilised at $\approx 6\,\mathrm{K}$.

Photoluminescence was separated from the pump with a long-pass filter.
Spectral analysis was carried out using a Czerny-Turner spectrometer (Andor Shamrock 750) coupled to a Peltier-cooled CCD camera (Andor iKon-M).
Mode tomography and hyperspectral imaging were obtained by mechanically scanning the corresponding image plane across the spectrometer entrance slit.
In the hyperspectral configuration, the back focal plane of the objective was imaged onto the detector, yielding angle- and energy-resolved far-field emission. In the energy-resolved tomography configuration, the sample surface (real-space image plane) was relayed to the slit to directly map spatial modes.

\subsection*{Reciprocal space map}
\label{sec:kspacePath}
The far-field (back-focal-plane) photoluminescence is recorded as a hyperspectral cube $I_{\mathrm{raw}}(p_x,p_y,E)$, where $(p_x,p_y)$ are the camera chip coordinates (pixel positions on the CCD) and $E$ the photon energy.
The pixel coordinates are converted to in-plane wave vectors $(k_\mathrm{x},k_\mathrm{y})$ by relating the size of the $k$-space sphere with the numerical aperture of the objective.
This process is explained in more detail in the Supplementary material~\ref{supp:pixel_to_k}.
Thus, we convert $I_{\mathrm{raw}}(p_x,p_y,E)\mapsto I(k_x,k_y,E)$.
We express the wave vector in units of the lattice by $k_y a/\pi\in[-1,1]$ (chain axis).

\subsubsection*{\textit{Zero and $\pi$ sublattice-phase cuts}}
\begin{figure}[h!]
\centering
\includegraphics[width=0.6\textwidth]{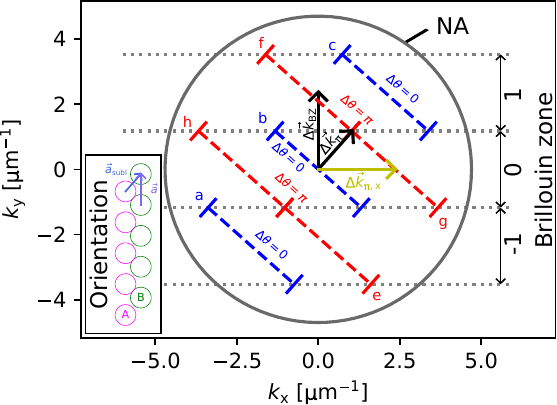}
\caption{   
\textbf{Reciprocal space cuts.}
Cuts through the far field of an SSH chain.
Optical access at large angles is limited by the objective numerical aperture.
The cuts corresponding to $\Delta \theta=0$ (zero effective sublattice phase shift) are displayed in blue, while the $\Delta \theta=\pi$ cuts are coloured red.
The Brillouin zones (up to second order) are highlighted with horizontal dashed lines.
A real-space sketch of the chain is included to show the connection between the vectors: $\Delta \vec{k}_\mathrm{BZ} \cdot \vec{a}=2\pi$ and $\Delta \vec{k}_\mathrm{\pi} \cdot \vec{a}_\mathrm{subl}=2\pi$.
The cuts labeled \textbf{a-c} ($\Delta \theta=0$) and \textbf{e-h} ($\Delta \theta=\pi$) are shown individually in Fig.\,\ref{fig:methods:indivCuts}.
}
\label{fig:methods:kSpaceCuts}
\end{figure}

\begin{figure}[h!]
\centering
\includegraphics[width=0.99\textwidth]{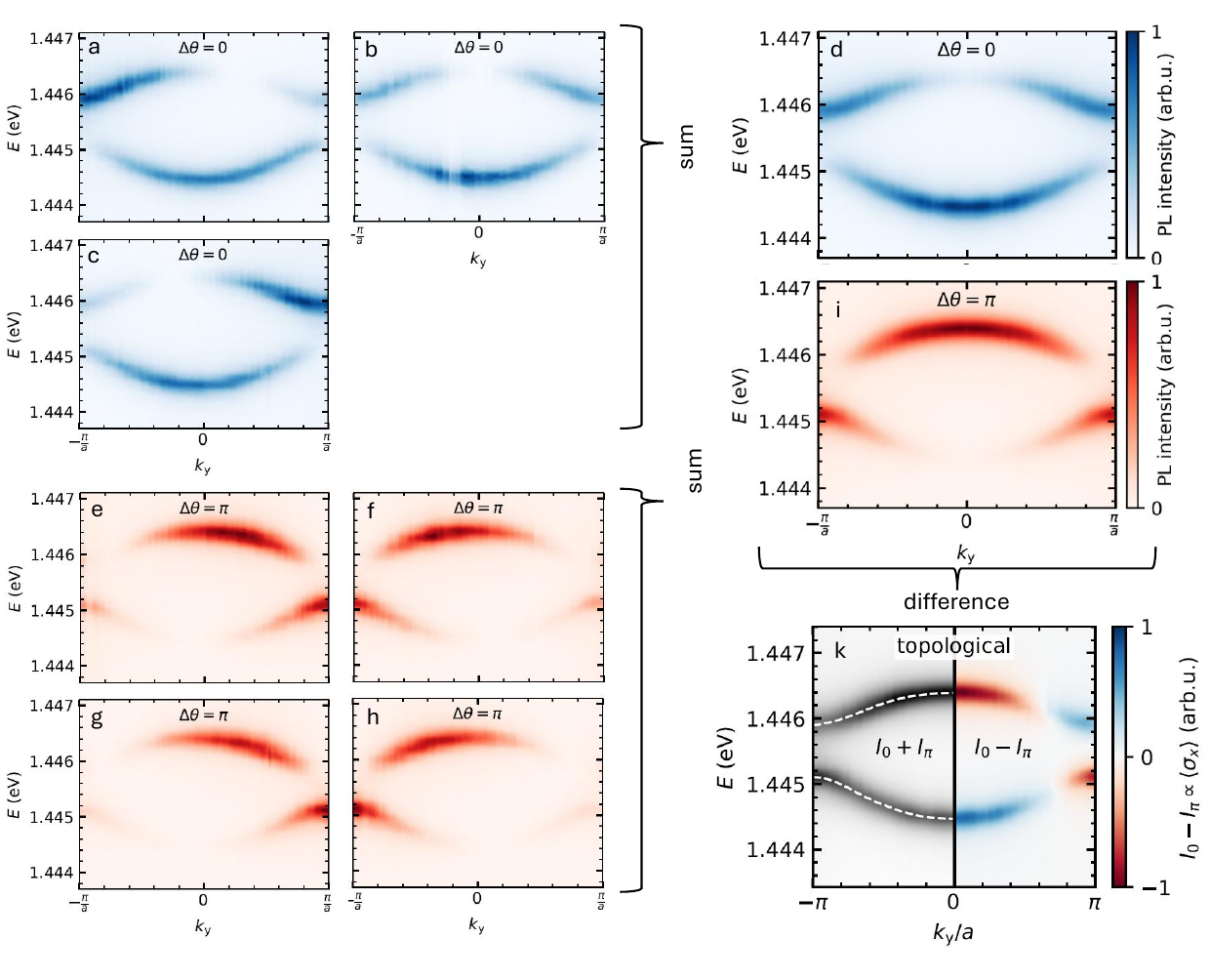}
\caption{
\textbf{Individual cuts of the topological configuration. a-c,}
Cuts through the far field along the $\Delta \theta = 0$ lines displayed in Fig.\,\ref{fig:methods:kSpaceCuts}. \textbf{d,} Sum of the three zero effective sublattice phase spectra. \textbf{e-h,} $\Delta \theta = \pi$ spectra displayed in Fig.\,\ref{fig:methods:kSpaceCuts}. \textbf{i,} Sum of the four $\pi$ effective sublattice phase spectra. \textbf{k,} Resulting difference spectrum (positive $k_\mathrm{y}$, showing the band inversion. The difference spectra presented in the main part are symmetrised with respect to $k_\mathrm{y}=0$.
}
\label{fig:methods:indivCuts}
\end{figure}

Let $\vec a_{\mathrm{subl}}$ denote the real-space vector connecting the sublattice centres
(defined in the main text). The effective sublattice phase shift $\Delta\theta$ is controlled by
choosing straight cuts in reciprocal space that are perpendicular to $\vec a_{\mathrm{subl}}$:
the ($\Delta\theta=0$)-family passes through $\vec k=\vec 0$, while the ($\Delta\theta=\pi$)-family is parallel-displaced by
\begin{align*}
\Delta\vec k_\pi=\frac{\pi}{|\vec a_{\mathrm{subl}}|}\cdot\frac{\vec a_{\mathrm{subl}}}{|\vec a_{\mathrm{subl}}|}.
\end{align*}
For convenience in plotting and to keep the Brillouin zone centred in each cut, we use an
equivalent purely perpendicular offset along $k_\mathrm{x}$,
\begin{align*}
\Delta \vec{k}_{\pi,\mathrm{x}}=\left ([\Delta\vec k_\pi]_\mathrm{x}+\frac{([\Delta\vec k_\pi]_\mathrm{y})^2}{[\Delta\vec k_\pi]_\mathrm{x}}, \;\; 0\right),
\end{align*}
where $[...]_\mathrm{x/y}$ is the $x$ or $y$-component of the vector.
Thus, both $\Delta\theta=0$ and $\Delta\theta=\pi$ cuts are centred around $k_\mathrm{y}=0$ while remaining perpendicular to $\vec a_{\mathrm{subl}}$.
For chains oriented along $y$, this offset is purely in $k_\mathrm{x}$.
The $k$-space cuts and the respective vectors are displayed in Fig.\,\ref{fig:methods:kSpaceCuts}.

\subsubsection*{\textit{Construction of $I_0$, $I_\pi$, and symmetrisation}}
Along each cut we sample the hyperspectral cube to obtain two $k_\mathrm{y}$-resolved spectra,
\begin{align*}
I_0(k_\mathrm{y},E)&\equiv I\big(\mathbf k\in\text{zero-phase cut},E\big),&
I_\pi(k_\mathrm{y},E)&\equiv I\big(\mathbf k\in\text{$\pi$-phase cut},E\big).
\end{align*}
We overlay all symmetry-related cuts available within the objective’s numerical aperture (NA) and average them to improve the signal-to-noise ratio as well as to overcome the parabolic density of states distribution of the unconfined polariton.
This $E\propto |k|^2$ maximum intensity paraboloid is inherently superimposed with the sublattice interference induced band intensity variation we are interested in. 
Finally, we symmetrise
$k_\mathrm{y}\!\to\!-k_\mathrm{y}$.

\subsubsection*{\textit{Observables}}
The sum and difference spectra
\begin{align*}
I_{\Sigma}(k_\mathrm{y},E)&=I_0(k_\mathrm{y},E)+I_\pi(k_\mathrm{y},E), &
I_{\Delta}(k_\mathrm{y},E)&=I_0(k_\mathrm{y},E)-I_\pi(k_\mathrm{y},E)
\end{align*}
are used throughout the work.
$I_{\Sigma}$ yields the structure-factor-corrected spectral weight (i.e. the pure band structure free of sublattice interference), while $I_{\Delta}$ is proportional to the sublattice symmetry $\langle\sigma_x\rangle(k_\mathrm{y},E)$ under the SSH conditions ($d_\mathrm{z}=0$ and $|\psi_\mathrm{A}|=|\psi_\mathrm{B}|$), thereby directly revealing the bulk band inversion at the Brillouin zone edge.
The same $k$-space cuts serve as inputs to the phase-fit procedure used
to reconstruct the Bloch eigenvectors and, hence, the effective Hamiltonian in the Pauli basis.
\subsubsection*{\textit{Relation to $\langle \sigma_\mathrm{x}\rangle$}}
The constructed difference spectra are directly related to the expectation value of the first Pauli matrix:
\begin{align*}
I_0 - I_\pi &\equiv|\psi_\mathrm{A}+\psi_\mathrm{B}|^2 - |\psi_\mathrm{A}-\psi_\mathrm{B}|^2\\
&=(\psi_\mathrm{A}+\psi_\mathrm{B})(\psi_\mathrm{A}^*+\psi_\mathrm{B}^*)-(\psi_\mathrm{A}-\psi_\mathrm{B})(\psi_\mathrm{A}^*-\psi_\mathrm{B}^*)\\
&=\psi_\mathrm{A}\psi_\mathrm{A}^*+\psi_\mathrm{A}\psi_\mathrm{B}^*+\psi_\mathrm{B}\psi_\mathrm{A}^*+\psi_\mathrm{B}\psi_\mathrm{B}^*-(\psi_\mathrm{A}\psi_\mathrm{A}^*-\psi_\mathrm{A}\psi_\mathrm{B}^*-\psi_\mathrm{B}\psi_\mathrm{A}^*+\psi_\mathrm{B}\psi_\mathrm{B}^*)\\
&=2\cdot(\psi_\mathrm{A}^*\psi_\mathrm{B}+\psi_\mathrm{B}^*\psi_\mathrm{A})\;.
\end{align*}
Thus, the difference spectra shown in this work are equivalent to the expectation value of the first Pauli matrix (for an infinitely extended chain), as detailed in equation\,\ref{eq:sigmaXBandInv}.
\clearpage

\subsection*{Individual cuts}

\subsection*{Hamiltonian and phase fit procedure}
\label{sec:phaseFitting}
Assuming an infinite chain, we reconstruct the Bloch eigenvectors and the effective $2\times2$ Hamiltonian from the far-field data by (i) locating the band energies $E_\pm(k_\mathrm{y})$, (ii) fitting the \emph{effective sublattice phase} dependence of the band intensities to extract the relative phase $\phi_\pm(k_\mathrm{y})$ between sublattices, and (iii) forming projectors to obtain $H(k_\mathrm{y})$ in the Pauli basis.

\subsubsection*{\textit{Band energies from $I_\Sigma=I_0+I_\pi$}}
From the reciprocal-space map (Methods: \emph{Reciprocal space map}) we build the zero- and $\pi$-phase spectra $I_0(k_\mathrm{y},E)$ and $I_\pi(k_\mathrm{y},E)$ along the perpendicular cuts.
Their sum
\begin{align*}
I_\Sigma(k_\mathrm{y},E)=I_0(k_\mathrm{y},E)+I_\pi(k_\mathrm{y},E)    
\end{align*}
suppresses sublattice interference and yields a structure-factor-corrected band map.
For each $k_\mathrm{y}$ we fit $I_{\Sigma, k_\mathrm{y}}(E)$ to a background plus two equal-height Lorentzians with centre energies $E_+$, $E_-$ and common FWHM $\Gamma$:
\begin{align*}
I_{\Sigma,k_\mathrm{y}}(E)\simeq A + \frac{B}{1+\bigl[2(E-E_{-,k_\mathrm{y}})/\Gamma\bigr]^2}+\frac{B}{1+\bigl[2(E-E_{+,k_\mathrm{y}})/\Gamma\bigr]^2}\,,
\end{align*}
thereby extracting the lower/upper band dispersions $E_-(k_\mathrm{y})$ and $E_+(k_\mathrm{y})$.
\subsubsection*{\textit{Phase sweep and cosine fit}}
To access the relative sublattice phase, we sample the intensity of each band while sweeping the \emph{effective} phase shift $\Delta\theta$ between sublattices.
More precisely, $\Delta\theta$ is controlled by translating the cut along the $\pi$-phase shift vector $\Delta\vec k_\mathrm{\pi,x}$ (Methods: \emph{Reciprocal space map}).
For each central point $\vec{k}_0(k_\mathrm{y})$ on the $\Delta\theta=0$ cut (line \textbf{b} in Fig.\,\ref{fig:methods:kSpaceCuts}), we interpolate the data along a line that is equally spaced in $\Delta\theta$:
\begin{align*}
 \vec{k}(\Delta\theta)= \vec{k}_0(k_\mathrm{y}) + \frac{\Delta\theta}{\pi}\,\Delta\vec k_{\pi,\mathrm{x}},\qquad \Delta\theta\in[\Delta\theta_{\min},\Delta\theta_{\max}]\,,
\end{align*}
with $\Delta\theta_{\min}=-1.2\pi$, $\Delta\theta_{\max}=+1.2\pi$.
At each $\vec{k}(\Delta\theta)$ we read out the far-field intensity at the fitted band energies,
\begin{align*}
I_\pm(k_\mathrm{y},\Delta\theta)=I\bigl(\mathbf k(\Delta\theta),E_\pm(k_\mathrm{y})\bigr),
\end{align*}
and normalise by the instantaneous total to remove variations due to the polariton effective mass (rotation-parabolically distributed maximum of intensity $E_\mathrm{max}\propto |k_{||}|^2$),
\begin{align*}
\tilde I_-(k_\mathrm{y},\Delta\theta)=\frac{I_-(k_\mathrm{y},\Delta\theta)}{I_-(k_\mathrm{y},\Delta\theta)+I_+(k_\mathrm{y},\Delta\theta)},\quad
\tilde I_+(k_\mathrm{y},\Delta\theta)=\frac{I_+(k_\mathrm{y},\Delta\theta)}{I_-(k_\mathrm{y},\Delta\theta)+I_+(k_\mathrm{y},\Delta\theta)}.
\end{align*}
To improve the signal-to-noise-ratio, we average over symmetry-equivalent points in reciprocal space by translating $\vec{k}(\Delta\theta)\mapsto \mathbf k(\Delta\theta)+2m\,\Delta\vec k_{\pi,\mathrm{tot}}+2n\,\mathbf k_{\mathrm{BZ}}$ for all integers $m,n\in\mathbb{Z}$ where the resulting line is within the microscope objective numerical aperture.
For each band at each $k_\mathrm{y}$ we then fit the $\Delta\theta$-dependence to a cosine with free offset and visibility,
\begin{align*}
\tilde I_\pm(k_\mathrm{y},\Delta\theta)\;\simeq A_\pm + \frac{B_\pm}{2}\Bigl(1+\cos\bigl(\Delta\theta-\phi_\pm(k_\mathrm{y})\bigr)\Bigr)\,,
\end{align*}
extracting the phase $\phi_\pm(k_\mathrm{y})$.
The fit returns $A\approx 0$ and $B\approx 1$ in the linear SSH regime, so the minima of $\tilde I_\pm$ approach zero.
This vanishing intensity directly underpins the equal-sublattice-localisation assumption $|\psi_\mathrm{A}|^2=|\psi_\mathrm{B}|^2=\tfrac{1}{2}$ (i.e. no staggered on-site energy $d_\mathrm{z}=0$), which maximises the interference fringe visibility.

\subsubsection*{\textit{Eigenvectors and Hamiltonian reconstruction}}
With $|\psi_\mathrm{A}|^2=|\psi_\mathrm{B}|^2=\tfrac{1}{2}$ fixed by symmetry, the Bloch eigenvectors at each $k_\mathrm{y}$ are completely determined by the fitted phases:
\begin{align*}
\ket{\psi_-(k_\mathrm{y})}=\frac{1}{\sqrt{2}}\!\begin{pmatrix}1\\ e^{i\phi_-(k_\mathrm{y})}\end{pmatrix},\qquad
\ket{\psi_+(k_\mathrm{y})}=\frac{1}{\sqrt{2}}\!\begin{pmatrix}1\\ e^{i\phi_+(k_\mathrm{y})}\end{pmatrix}.
\end{align*}
We enforce orthogonality by setting $\phi_-(k_\mathrm{y})=\phi_+(k_\mathrm{y})+\pi$ (consistent with chiral symmetry).
The spectral decomposition
\begin{align*}
H(k_\mathrm{y})\;=\;E_-(k_\mathrm{y})\,\ket{\psi_-(k_\mathrm{y})}\!\bra{\psi_-(k_\mathrm{y})}\;+\;E_+(k_\mathrm{y})\,\ket{\psi_+(k_\mathrm{y})}\!\bra{\psi_+(k_\mathrm{y})}
\end{align*}
is then evaluated and projected onto the Pauli basis to obtain
\begin{align*}
d_0(k_\mathrm{y})&=\tfrac{1}{2}\,\mathrm{Tr}\,H,\qquad
d_\mathrm{x}(k_\mathrm{y})=\tfrac{1}{2}\,\mathrm{Tr}\bigl(H\,\sigma_x\bigr),\\
d_\mathrm{y}(k_\mathrm{y})&=\tfrac{1}{2}\,\mathrm{Tr}\bigl(H\,\sigma_\mathrm{y}\bigr),\qquad
d_\mathrm{z}(k_\mathrm{y})=\tfrac{1}{2}\,\mathrm{Tr}\bigl(H\,\sigma_\mathrm{z}\bigr)\approx 0,
\end{align*}
from which we plot the $(d_\mathrm{x},d_\mathrm{y})$ trajectory and evaluate the winding.
In all datasets $d_\mathrm{z}(k_\mathrm{y})$ remains negligible compared to $\sqrt{d_\mathrm{x}^2+d_\mathrm{y}^2}$, consistent with the SSH symmetry and the observed near-zero minima in the phase sweeps.

\clearpage
\renewcommand{\thesection}{S\arabic{section}}
\renewcommand{\thefigure}{S\arabic{figure}}
\renewcommand{\theequation}{S\arabic{equation}}
\renewcommand{\thetable}{S\arabic{table}}
\setcounter{section}{0}
\setcounter{figure}{0}
\setcounter{equation}{0}
\setcounter{table}{0}

\section*{Supplementary Information}\label{supplementary}

\section{Connection between winding number and band inversion}
\label{supp:section1}
We experimentally observe a band inversion that is linked to a non-trivial winding number $\nu$.
While it is commonly known that $|v|<|w| \rightarrow \nu=1$ and $|v|>|w| \rightarrow \nu=0$, the band inversion associated with the topology of the SSH model is thus far unreported.
Therefore, we show in the following that an inverted band structure in $\langle\sigma_\mathrm{x}\rangle$ is directly tied to a non-trivial winding number.
The Hamiltonian of the SSH model at $k_\mathrm{y}=0$, $\hat{H}_0$, with eigenvectors $\hat{H}_0\vec{\psi}_{0,1\!/2}=E_{0,1\!/2}\vec{\psi}_{0,1\!/2}$ is given by
\begin{align}
    \hat{H}_0 = \begin{pmatrix}
0 & v + w \\
v + w & 0
\end{pmatrix} ,\;\;\;
E_{0,1\!/2} =\pm(v+w),\;\;\;
\vec{\psi}_{0,1\!/2}=\begin{pmatrix}
1  \\
\pm1
\end{pmatrix},
\label{eq:hk0}
\end{align}
while at the edge of the Brillouin zone $k_\mathrm{y}=\frac{\pi}{a}$
\begin{align}
    \hat{H}_\pi = \begin{pmatrix}
0 & v - w \\
v - w & 0
\end{pmatrix} ,\;\;\;
E_{\pi,1\!/2} =\pm(w-v),\;\;\;
\vec{\psi}_{\pi,1\!/2}=\begin{pmatrix}
\mp1  \\
1
\end{pmatrix}.
\label{eq:hkpi}
\end{align}
Thus, for positive hopping $0<w,v$, the negative eigenvalue at $k_\mathrm{y}=0$ is always $E_\mathrm{0,-}=-(v+w)$ with $\vec{\psi}_{0,-}=(1,1)^T$, while at $k_\mathrm{y}=\frac{\pi}{a}$, it is given by $E_\mathrm{\pi,neg}=-|w-v|$ with the eigenvector $\vec{\psi}_{\pi,-}=(\mathrm{sgn}(v-w),1)^T$.
The sublattice symmetry must therefore invert when going from $k_\mathrm{y}=0$ to $k_\mathrm{y}=\frac{\pi}{a}$, when $\mathrm{sgn}(v-w)=-1 \Leftrightarrow v<w$, linking the band inversion to the non-trivial winding number~\cite{Asboth2016}.

\clearpage

\section{Converting CCD pixel coordinates to in-plane-$k$}
\label{supp:pixel_to_k}
While for many studies it is sufficient to approximate a linear relationship between camera pixel coordinates $(p_\mathrm{x},p_\mathrm{y})$ and in-plane-$k$-vector (especially for small numerical apertures), our work requires a more precise treatment.
This process is explained in the following.

\subsubsection*{\textit{Converting pixels to angle}}
\begin{figure}[h!]
\centering
\includegraphics[width=1\textwidth]{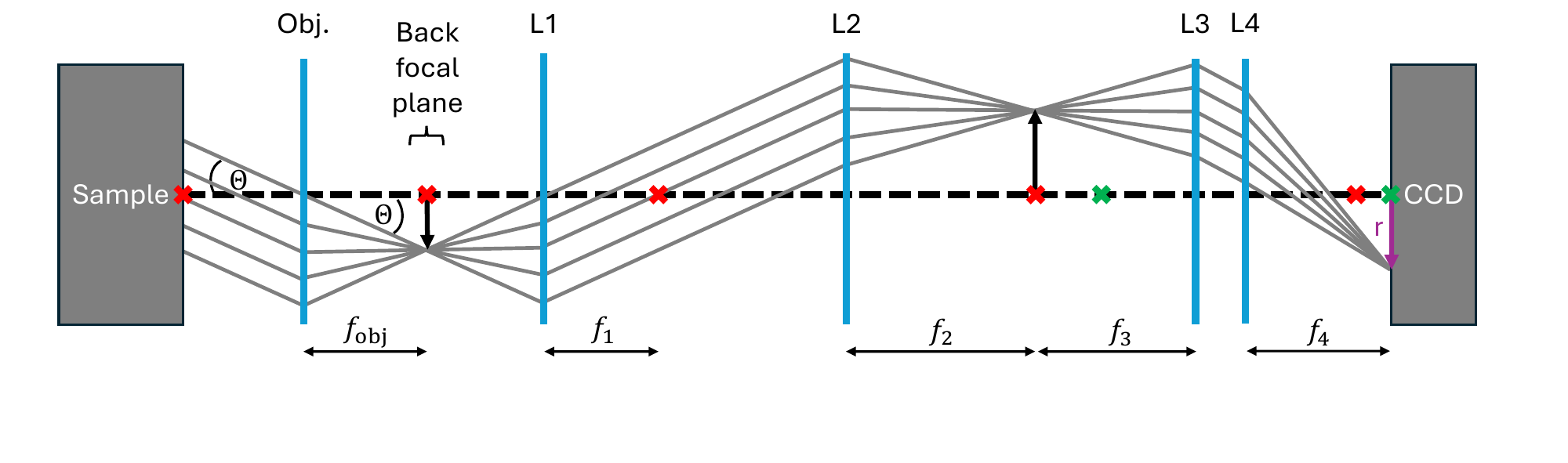}
\caption{   
\textbf{Detection path in reciprocal space.} Optical path of light that is emitted by the sample under an angle $\Theta$.
The light is projected onto one position on the CCD (detector) at radius $r$. 
}
\label{fig:supp:detectionPath}
\end{figure}

Due to the rotation symmetry of the detection path (objective and lenses) around the optical axis, the conversion process can be reduced to the treatment of the angle of the emitted light with respect to the optical axis, in relation to the radial distance on the CCD $r=\sqrt{p_\mathrm{x}^2+p_\mathrm{y}^2}$, where $(p_\mathrm{x},p_\mathrm{y})=(0,0)$ is the intersecting point between the optical axis and the CCD.
Therefore, polar coordinates offer a suitable coordinate system.
A sketch of the detection path is shown in Fig.\,\ref{fig:supp:detectionPath}, including the occurring focal lengths.
The figure illustrates the projection of one specific emission angle into one CCD position.
With the help of trigonometry and considering all focal lengths, the emission angle is given by
\begin{align*}
\Theta = \arctan \left( \frac{r}{f_\mathrm{obj}} \frac{f_3}{f_4}\frac{f_1}{f_2}\right) \; .
\end{align*}
In this process, the angle of the polar coordinates in the CCD plane $\phi=\arg(p_\mathrm{x}+ i p_\mathrm{y})$ -- not to be confused with $\Theta$ -- is conserved due to the rotational symmetry, when translating it to the angle of the emission direction in 3D space.
Thus, the $x$ and $y$ emission angles from the sample plane can be calculated as follows:
\begin{align*}
    \Theta_\mathrm{x}=\Theta \cdot \cos(\phi), \;\; \Theta_\mathrm{y}=\Theta \cdot \sin(\phi) \; .
\end{align*}
In this step, a rotational tilt of the sample with respect to the objective has to be taken into account in the form of small angle offsets $\Theta_\mathrm{x/y}\rightarrow\Theta_\mathrm{x/y} + \delta \Theta_\mathrm{x/y}$ before converting back to absolute emission angle $\Theta_\mathrm{corr}=\sqrt{(\Theta_\mathrm{x}+\delta \Theta_\mathrm{x})^2 + (\Theta_\mathrm{y}+\delta \Theta_\mathrm{y})^2}$.
\\
Note, that an independent treatment of the $p_x$ and $p_y$ coordinate on the CCD is not correct (or only for small angles), due to the nonlinearity of the conversion equation.

\subsubsection*{\textit{Converting angle to $k$}}
In a microcavity, the total wave vector is given by the sum of the wave vector perpendicular to the cavity $\vec{k}_\perp=(0, 0, k_\perp)$ and the in-plane $k$-vector $\vec{k}_\parallel=(k_\mathrm{x}, k_\mathrm{y}, 0)$.
Using $E_\mathrm{ph}=\hbar c |k|$, the energy is given by
\begin{align*}
    E=\hbar c \sqrt{k_\mathrm{x}^2+k_\mathrm{y}^2+k_\perp^2}
     \equiv\hbar c \sqrt{k_\mathrm{\parallel}^2+k_\perp^2} \; .
\end{align*}
The contribution perpendicular to the cavity is fixed and given by the cavity wavelength $k_\perp=\frac{2 \pi n_\mathrm{cav}}{\lambda_\mathrm{cav}}$ and its refractive index $n_\mathrm{cav}$~\cite{Deng10}.
Due to the cavity refractive index, the angle of light emitted into the detection path $\Theta$ is modified inside of the cavity according to Snell's law
\begin{align*}
n_\mathrm{cav} \sin \Theta_\mathrm{cav}&=\sin\Theta_\mathrm{corr} \\
\Leftrightarrow \Theta_\mathrm{cav} &= \arcsin \left( \frac{\sin\Theta_\mathrm{corr}}{n_\mathrm{cav}} \right)
\end{align*}
Furthermore, due to the orthogonality $\vec{k}_\perp \perp \vec{k}_\parallel$, the relation $\tan \Theta_\mathrm{cav}=\frac{k_\mathrm{\parallel}}{k_\perp}$ is given.
Using these relations, the in-plane wave vector can be calculated from the emission angle using~\cite{Deng10}
\begin{align*}
k_\parallel = \frac{2 \pi n_\mathrm{cav}}{\lambda_\mathrm{cav}} \cdot \tan\left [\arcsin \left( \frac{\sin\Theta_\mathrm{corr}}{n_\mathrm{cav}}\right)\right] \; .
\end{align*}
While this equation holds for planar microcavities, the cavity refractive index has to be set to unity $n_\mathrm{cav}=1$, because the lattice periodicity is defined in terms of vacuum wavelength.
Otherwise, the band structure will not be periodic in $k$ (higher order Brillouin zones), impeding backfolding procedures -- especially with low-linewidth samples.
Note, that the absolute values of emission angle and in-plane wave vector have to be used again due to the nonlinearity of the conversion formula at larger angles (high numerical apertures).
Thus, the $x$ and $y$ components of the in-plane wave vector can be determined using the angle $\phi$ of the corresponding pixel in polar coordinates again:
\begin{align*}
k_\mathrm{x}=k_\parallel\cdot\cos\phi, \;\; k_\mathrm{y}=k_\parallel\cdot\sin\phi \; .
\end{align*}
\clearpage

\end{document}